\documentclass[12pt]{article}
\usepackage{epsf}
\title{Pentaquark decay is suppressed by chirality conservation}

\author{B.L.Ioffe and A.G.Oganesian\\
Institute of Theoretical and Experimental Physics,\\
B.Cheremushkinskaya 25, 117218 Moscow,Russia}
\date{}
\begin{document}
\pagestyle{empty}

\maketitle

\newcommand{\be}{\begin{equation}}
\newcommand{\ee}{\end{equation}}

\def\la{\mathrel{\mathpalette\fun <}}
\def\ga{\mathrel{\mathpalette\fun >}}
\def\fun#1#2{\lower3.6pt\vbox{\baselineskip0pt\lineskip.9pt
\ialign{$\mathsurround=0pt#1\hfil##\hfil$\crcr#2\crcr\sim\crcr}}}

\vspace{1cm}

It is shown, that if the pentaquark $\Theta^+ = uudd\bar{s}$
baryon can be represented by the local quark current
$\eta_{\Theta}$, its decay $\Theta^+ \to n K^+ (p K^0)$ is
forbidden in the limit of chirality conservation. The
$\Theta^+$decay width $\Gamma$ is proportional to $\alpha^2_s
\langle 0 \vert \bar{q} q \vert 0 \rangle^2$, where $\langle 0
\vert \bar{q} q \vert 0 \rangle$, $q = u,d,s$ is quark condensate,
and, therefore, is strongly suppressed. Also the polarization
operator of the pentaquark current with isospin 1 is calculated
using the operator product expansion and estimation for it mass is
obtained .
%\end{abstract}


\vspace{1cm}

\normalsize

 Last year, the exotic baryon resonance $\Theta^+$ with
quark content $\Theta^+ = uudd\bar{s}$ and mass 1.54 GeV [1,2] had
been discovered. Later, the existence of this resonance was
confirmed by many other groups, although some searches for it were
unsuccessful. (see [3] for the review). $\Theta^+$ baryon was
predicted in 1997 by D.Diakonov, V.Petrov and M.Polyakov [4] in
the Chiral Soliton Model as a member of antidecouplet with
hypercharge $Y = 2$. The recent theoretical reviews are given in
[5,6]. $\Theta^+$ was observed as a resonance in the systems
$nK^+$ and $pK^0$. No enhancement was found in $pK^+$ mass
distributions, what indicates on isospin $T=0$ of $\Theta^+$ in
accord with theoretical predictions [4].

One of the most interesting features of $\Theta^+$ is its very
narrow width. Experimentally, only an upper limit was found, the
stringer bound was presented in [2]: $\Gamma < 9 MeV$. The phase
analysis of $KN$ scattering results in the even stronger limit on
$\Gamma$ [7], $\Gamma < 1 MeV$. A close to the latter limitation
was found in [8] from the analysis of $Kd \to ppK$ reaction and in
[9] from $K+Xe$ collisions data [2]. The Chiral Quark Soliton
Model gives the estimation [4]: $\Gamma_{CQSM} \la 15 MeV$
(R.E.Jaffe [10] claims that this estimation has a numerical error
and in fact $\Gamma_{CQSM}\la 30 MeV$ -- see, however, [11]). In
any way, the estimation [4] for $\Gamma_{CQSM}$ follows from the
cancellation of large and uncertain numbers and it is not quite
reliable. Therefore, till now the narrow $\Theta^+$ width is a
theoretical puzzle.

We suggest here its qualitative explanation. Suppose, that
$\Theta^+$ may be represented by the local 5 quark current
$\eta_{\Theta}^{T=0}$ corresponding to isospin $T=0$. An example
of such current is:

%1
\be
 \eta^{T=0}_{\Theta} = \{\varepsilon^{abc} [(d^a C \gamma_{\mu}
d^b) u^c - (d^a C \gamma_{\mu} u^b) d^c ] \cdot \bar{s}
\gamma_{\mu} \gamma_5 u + u\leftrightarrow d \}/\sqrt2  \ee where
$a,b,c$ are color indexes , $C$ - charge conjugation matrix,
$u,d,s$ -- are quark fields. Suppose also, that the amplitude of
$\Theta^+ \to nK^+$ decay is proportional to vacuum average

%2
\be
{\cal{M}} = \langle 0 \vert T \{\eta_n(x), j ^{\lambda}_5(y),~
\bar{\eta}_{\Theta}(0)\} \vert 0 \rangle \ee where $\eta_n(x)$ is
the neutron quark current [12]
%3

$$\eta_n =\varepsilon^{abc} (d^a C \gamma_{\mu} d^b) \gamma_5
\gamma_{\mu} u^c $$ and $j_{\mu 5} = \bar{s} \gamma_{\mu}\gamma_5
u$ is the strange axial current. Let us neglect quark masses and
perform the chiral transformation $q \to \gamma_5 q$. It is
evident, that $\eta_n$ and $j_{\mu 5}$ are even under such
transformation, while $\eta_{\Theta}$ is odd. Therefore, the
matrix element (2) vanishes in the limit of chiral symmetry. It is
easy to see, that this statement is valid for any form of
pentaquark and nucleon quark currents (spinless and with no
derivatives). In the real world the chiral symmetry is
spontaneously broken. The lowest dimension operator, corresponding
to violation of chiral symmetry is $\bar{q}q$. So, the matrix
element (2) is proportional to quark condensate $\langle 0 \vert
\bar{q} q \vert 0 \rangle$. Moreover, if $\Theta^+$ is a genuine
5-quark state (not, say, the $NK$ bound state), then in (2) the
hard gluon exchange is necessary, what leads to additional factor
of $\alpha_s$. The necessity to have gluonic exchange  in order to
get nonvanishing value of ${\cal{M}}$ is confirmed by direct
calculation of ${\cal{M}}$ for any $\eta_{\Theta}$ by the QCD sum
rules (s.r.) method for three point function suggested in  [13].
We come to the conclusion, that $\Gamma_{\Theta} \sim \alpha^2_s
\langle 0 \vert \bar{q} q \vert 0 \rangle^2$, i.e.,
$\Gamma_{\Theta}$ is strongly suppressed. This conclusion takes
place for any genuine 5-quark states -- the states formed from 5
current quarks at small separation, but not for potentially
bounded $NK$-resonances, corresponding to large relative
distances. There are no such suppression for the latters.

Now let us to discuss the calculation of two-point correlator. The
calculation of two point correlator for the $\eta_{\Theta}^{T=0}$
have some problem and results will be presented elsewhere.
Consider now the pentaquark current corresponds to the isospin 1

\be
\eta_{\Theta}^{T=1}(x) = [\varepsilon^{abc} (d^a C \sigma_{\mu
\nu} d^b) \gamma_{\nu} u^c \cdot \bar{s} \gamma_{\mu} \gamma_5 u -
(u \leftrightarrow d)]/\sqrt{2}, \ee
 As can
be easily seen, all considerations, performed above, are valid for
$\eta^{T=1}_{\Theta}$ too as for any other.

Let us now calculate in QCD the polarization operator
%4
\be
\Pi(p) = i \int d^4 x e^{ipx} \langle 0 \vert T
{\eta_{\Theta}(x),~ \bar{\eta}_{\Theta} (0)} \vert 0 \rangle \ee
where $\eta_{\Theta}^{T=1}$ is given by (3) and examine, if that
it may be represented by the contribution of pentaquark state with
$T=1$ and exited states (continuum). Consider $p^2 < 0$ and $\vert
p^2 \vert$ large enough, use the operator product expansion (OPE)
and QCD s.r.method for baryons [12]. The Lorenz structure of
$\Pi(p)$ has the form

%5
\be
\Pi(p) = \hat{p} \Pi_1 (p^2) + \Pi_2(p^2) \ee $\Pi_1(p)$ is
calculated with the account of operators up to dimension 12,
$\Pi_2(p)$ -- up to dimension 13. Masses of $u$ and $d$- quarks
are neglected, the $s$-quark mass $m_s$ is accounted in the first
order. Factorization hypothesis is assumed for operators of higher
dimensions, operators anomalous dimensions are neglected, as well
as $\alpha_s$ corrections. On the other side, represent $\Pi(p)$
in terms of physical states contributions -- $\Theta^{T=1}$ and
continuum, starting from some $(p^2)_0 \equiv s_0$. After Borel
transformation the sum rules are given by

%6
$$ \frac{3M^{12}}{560} \left \{\frac{7}{6} E_5 + (\frac{7}{18}
b-\frac{28}{3} m_s a)\frac{E_3}{M^4}-\frac{35}{18}[(5+8
\gamma)a^2-\frac{31}{4}m_s m^2_0 a]\frac{E_2}{M^6}+ \right.$$
$$\left.+ \frac{35}{6} (\frac{5}{2}+\frac{13}{3} \gamma) m^2_0 a^2
\frac{E_1}{M^8} + \Biggl [\frac{560}{9} m_s a^3 - \frac{175}{48}
(1+2 \gamma) m^4_0 a^2 \right.$$
\be
\left.-\frac{35}{54} (19 \gamma - \frac{3}{8}) b a^2 \Biggr ]
\frac{E_0}{M^{10}} + \frac{700}{27} (1.6 \gamma-0.2)
\frac{a^4}{M^{12}} \right \}=\bar{\lambda}^2 e^{-m^2/M^2} \ee

$$ \frac {M^{10}}{32} a \left \{\frac{m_s}{10} \frac{M^2}{a}E_5
-\frac{1}{5}(8-\gamma)E_4+\frac{1}{12}(29-3 \gamma)m^2_0
\frac{E_3}{M^2}-\Biggl [(\frac{9}{4}-\frac{17}{36}
\gamma)b-\frac{20}{3} m_s a \Biggr ]\frac{E_2}{M^4}+ \right.$$
 $$\left.+ \Biggl [8(\frac{4}{3} +
\frac{5}{6} \gamma) a^2-\frac{65}{6}m_sm^2_0 a \Biggr ]
\frac{E_1}{M^6} - (14+\frac{25}{3} \gamma)m^2_0 a^2
\frac{E_0}{M^8}+ \Biggl [(3+\frac{131}{288}+\frac{37}{54}
\gamma)m^4_0 a^2- \right.$$
\be
\left.\frac{112}{9} m_s a^3 \Biggr ] \cdot \frac{1}{M^{10}} \right
\} = \bar{\lambda}^2 me^{-m^2/M^2},\ee

where $m$ is the $\Theta^{T=1}$-mass, $M$ is the Borel parameter,
$(q=u,d)$
%8
$$ a = -(2 \pi)^2 \langle 0 \vert \bar{q} q \vert 0 \rangle, ~
\gamma = \langle 0 \vert \bar{s}s \vert 0 \rangle /\langle 0 \vert
\bar{q} q \vert 0 \rangle $$
 $$b = (2 \pi)^2 \langle 0 \vert \frac{\alpha_s}{\pi} G^2 \vert 0
 \rangle$$
 \be
 g \langle 0 \vert \bar{q} \sigma_{\mu \nu}
 (\lambda^n/2) G^n_{\mu \nu} q \vert 0 \rangle  \equiv m^2_0 \langle
 0 \vert \bar{q} q \vert 0 \rangle
 \ee
 and the sign of $g$ is defined by the form of covariant
 derivative $\nabla_{\mu} = \partial_{\mu} - ig(\lambda^n/2)
 A^n_{\mu}$. $\bar{\lambda}$ is given by the matrix element
 %9
 \be
 \langle 0 \vert \eta_{\Theta} \vert \Theta \rangle = \lambda
 \upsilon_{\Theta}
 \ee
 where $\upsilon_{\Theta}$ is $\Theta^{T=1}$ spinor, $\bar{\lambda}^2
 = (2 \pi)^8 \lambda^2$. Continuum contributions are transferred to the left-hand
 sides (l.h.s) of the s.r. resulting in appearance  of the factors
 %10
 \be
 E_n(\frac{s_0}{M^2}) = \frac{1}{n!} \int\limits^{s_0/M^2}_{0} dz
 z^n e^{-z}
 \ee

%\vspace{5cm}

\begin{figure}
\epsfxsize=10cm \epsfbox{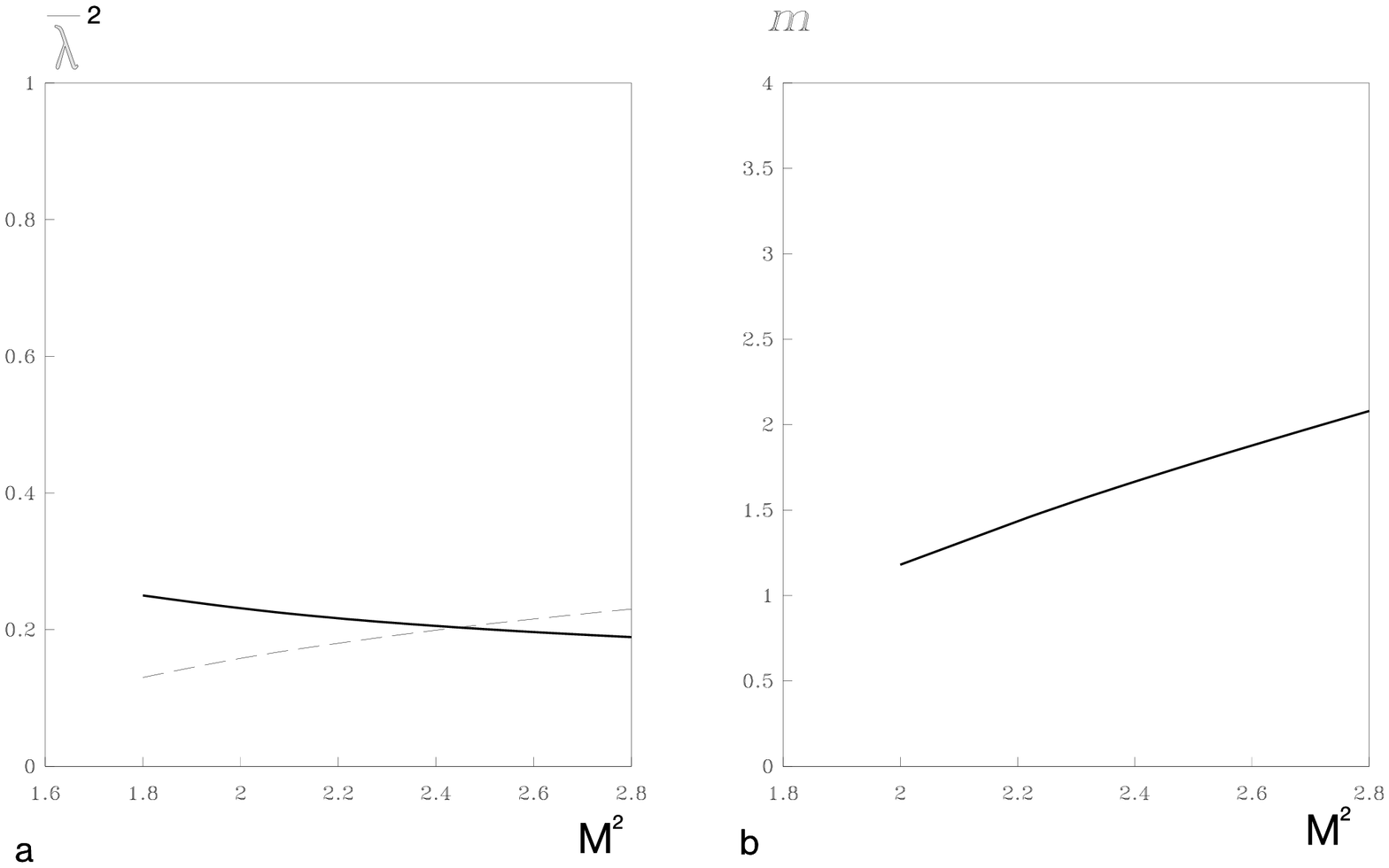} \caption{The $M^2$ dependence of
the sum rules (6),(7): a) $\bar{\lambda}^2$ from eq.(6) -- solid
line, $\bar{\lambda}^2$ from eq.(7) -- dashed line; b) $m$
obtained as a ratio of (7) to (6).  }
\end{figure}

The values of $\bar{\lambda}$, determined from eqs.(6) and (7)
 are plotted in Fig.1 (Fig.1a, for $m$ the  value of
 $m_{\Theta}^{T=1}=1.7 GeV$ was put in), and the value of $m$ obtained
 as a ratio of (7) to (6). The parameters were taken in accord
 with the recent determination of QCD condensates [14,15]  at
 normalization point $\mu^2=2 GeV^2$: $a=0.63 GeV^3$, $b=0.24
 GeV^4$, $m^2_0=1 GeV^2$, $m_s=0.15 GeV$, $\gamma=0.8$. It was
 chosen $s_0=4.5 GeV^2$.

 Few remarks are in order. The first term in (3)
 contains two left $d_Ld_L$ or two right $d_Rd_R$ quark components,
  while the neutron current $\eta_n$ is
 proportional to $d_Ld_R$ (see [12], eq.61). Therefore, in the
 chiral limit two-hadron reducible contributions [16] are absent
 in the case of the $\eta_{\Theta}^{T=1}$ current (3). The inspection
 of the s.r.'s shows, that the main contributions arise from
 operators of high dimensions (d=6,8 in (6) and d=5,9,11 in (7)),
 unlike the case of normal hadrons,
 where low dimension operators are dominant. This means, that
 pentaquark indeed differs very much from usual hadrons. There is
 a remarkable cancellation in (6) and (7) among the contributions
 of various operators. In consequence of this cancellation,
 the accuracy of the sum rules strongly depends on
 accuracy of the parameters, (especially $m_0^2$, but not only). Therefore we estimate
 the accuracy of sum rules (6,7) not better then $25\%$. For this
 reason as well because  not quite good agreement of two curves on Fig.1.
 basing on our calculations it is hard to insist on
  the existence of  the pentaquark state with isospin 1 .

Nevertheless, if one want to use this sum rules to obtain
estimation of mass for  pentaquark state with isospin 1, then the
value of $m_{\Theta}$ may be estimated as
 $m_{\Theta}^{T=1} = 1.7 \pm 0.4 GeV$ (see Fig.1b) The positiveness of the l.h.s of
 (7) clearly shows that the parity of $\Theta^{T=1}$ is positive.
 The result only slightly varies at the variation of $s_0$ within
 $10-15\%$.

 The QCD s.r.calculations of pentaquark masses with local
 $\eta_{\Theta}$ were performed in
 [17-19]. Unfortunately, nonsuitable chirally nonvariant 5-quark
 currents were chosen and the results  change drastically after
 subtraction of two-hadron reducible contributions [16]. And
 besides, in [17] only one structure was considered and important
 terms of OPE were omitted.

 Consider now the current $\eta_{\theta}$ similar to (3) but where persist the sum
 of two terms, instead of the difference in (3).
 This current is a mixture of two isospin states, $T=0$
 and $T=2$. The calculation show, that the s.r.(7) is
 essentially smaller than (6). So, in this case, there is no
 resonance  structure at the masses $1.5-2.0 GeV$, only a
 background more or less equally populated by the states of
 positive and negative parities (at the total angular momentum
 $j=1/2$).

The authors are thankful to M.Nielsen and to N.Kotchelev and
H.-J.Lee , who had found the errors in the initial version of the
paper.

One of the authors (A.Oganesian) is indebted to K.Goeke and to
M.Polyakov for their hospitality In Bochum university and for
useful discussions.

This work was supported in part by INTAS grant 2000-587, by RFBR
grant 03-02-16209 and by funds from EC to the project "Study of
Strongly Interacting Matter" under contract 2004
No.R113-CT-2004-506078.

\end{document}